\documentclass{PoS}

\usepackage{caption}
\usepackage{subcaption}
\usepackage{cite}

\usepackage{amsmath}
\def\colour4colour#1{\Blue{#1}}

\newcommand{\as}{a_{\rm s}}
\newcommand{\btz}{\beta_0}

\newcommand{\hspn}{{\hspace{-4mm}}}

\newcommand{\beq}{\begin{equation}}
\newcommand{\eeq}{\end{equation}}
\newcommand{\bea}{\begin{eqnarray}}
\newcommand{\eea}{\end{eqnarray}}

\newcommand{\als}{\alpha_{\rm s}}

\def\a#1{{a_{\rm s}^{\:\!#1}}}
\def\al#1{{\alpha_{\rm s}^{\:\!#1}}}

\def\nc{{n_c}}
\def\ncs{{n_{c}^{\,2}}}
\def\nct{{n_{c}^{\,3}}}

\def\cf{{C^{}_F}}

\def\nf{{n^{}_{\! f}}}

\def\nfs{{n^{\,2}_{\! f}}}
\def\nft{{n^{\,3}_{\! f}}}

\def\fl11{fl_{11}}

\title{
\vspace*{-2.3cm}
\begin{minipage}{\textwidth}
{\normalfont\small DESY 19-104, LTH 1212
\hspace{\fill} August 2019}\\
\end{minipage}\\[42pt]
 Soft corrections to inclusive DIS at four loops and beyond}

\ShortTitle{Soft corrections to inclusive DIS}

\author{\speaker{Goutam Das \phantom {g\hspace*{-3mm}}}\\
        Theory Group, Deutsches Elektronen-Synchrotron (DESY), D-22607 Hamburg, Germany\\
        E-mail: \email{goutam.das@desy.de}}
        
\author{Sven-Olaf Moch \phantom {g} \\
        II. Institute for Theoretical Physics, Hamburg University, D-22761 Hamburg, Germany\\
        E-mail: \email{sven-olaf.moch@desy.de}}

\author{Andreas Vogt\\
        \mbox{Department of Mathematical Sciences, University of Liverpool, Liverpool L69 3BX, 
        United~Kingdom}\\
        E-mail: \email{andreas.vogt@liverpool.ac.uk}
        \\ \\ }

\abstract{
We study the threshold corrections to the structure functions in  
deep-inelastic scattering (DIS) at the fifth logarithmic (N$^4$LL) order
of the soft-gluon exponentiation in massless perturbative~QCD. 
Using recent results for the splitting functions and the quark 
form factor, we derive the fourth-order contribution to the coefficient 
$f^{\,\rm q}$ of the form factor and from it the N$^4$LL part of the 
exponentiation coefficient $B^{\,\rm DIS}$ in the limit of a large number 
of colours.
An approximation scheme is shown that leads to sufficiently accurate 
N$^4$LL results for full QCD. The N$^4$LL corrections are small and lead 
to a further stabilization of the perturbative expansion for the 
soft-gluon exponent.
}

\FullConference{XXVII International Workshop on Deep-Inelastic Scattering and Related Subjects - DIS2019\\
		8-12 April, 2019, Torino, Italy}

\begin{document}

\section{Introduction}

\noindent
The Wilson coefficients (coefficient functions) for the structure functions 
of inclusive DIS have been a subject of research since the early days of QCD. 
These quantities are not only relevant for determining the parton distribution 
functions (PDFs) and the strong coupling constant $\als$ using structure
function data, see, e.g., \cite{Accardi:2016ndt}, 
but also to other processes and less inclusive observables in DIS, see, e.g., 
\cite{ApplCdis}.
The main Wilson coefficients for DIS are presently known up to the third order
in $\als$ in massless perturbative QCD \cite{Cdis3}.
Their perturbative expansion is well-behaved except close to the kinematic
endpoints $x = 0$ and $x = 1$ of the Bjorken variable.
The dominant terms \mbox{$\ln^{\,\ell}(1-x)/(1-x)_+$} in the latter 
(threshold) limit are resummed by the soft-gluon exponentiation, see, e.g.,  
\cite{Catani:1996yz,Contopanagos:1996nh,Catani:1998tm}, 
which is best formulated in Mellin $N$-space \cite{Catani:1996yz}. 
So far this resummation has been performed up to the next-to-next-to-next-to-%
leading logarithmic (N$^3$LL) accuracy \cite{Moch:2005ba}.

\vspace*{0.5mm}
The threshold resummation coefficients are closely related to the large-$x$
limit of the quark-quark splitting functions for the PDFs and to the quark
form factor \cite{ResFact,Ravindran:2006cg}
which are both fully known to order $\al3$ 
\cite{Pnnlo,Moch:2005id,Moch:2005tm,FF3}.
Recently the computations of these quantities have been extended to order 
$\al4$ in the (L$n_c$) limit of a large number of colours
\cite{Lee:2016ixa,Moch:2017uml}. 
Together with approximate results for the $\nf$-independent \cite{Moch:2018wjh}
and exact expression for the $\nf$-dependent contributions to the cusp 
anomalous dimension in full QCD  \cite{A4nf}
these results facilitate the effective extension of the threshold resummation
for the DIS Wilson coefficients to the next (N$^4$LL) logarithmic order.
In~the following we recall the theoretical framework, present the N$^4$LL 
resummation coefficient and briefly address the numerical implications of 
this result for the resummation of DIS in~QCD.

\section{Theoretical framework and new fourth-order coefficients}

\noindent
The all-order large-$N$ behaviour of the DIS Wilson coefficients for $F_1$, 
$F_2$ and $F_3$ can be written~as 
\beq
\label{eq:cNres}
  C^{\,N}(Q^2) \: =\: 
  g_{0}^{}(Q^2) \cdot \exp\, [G^{\,N}(Q^2)]   
  \: + \: {\cal O}(N^{-1}\ln^{\:\!n} N) \; ,
\eeq
where the resummation exponent $G^{\,N}$ of the dominant $N^{0}\ln^{\:\!n} N$ 
contributions is given by \cite{Ddis0}
\begin{align}
\label{eq:GNint}
  G^{\,N}=\int_0^1 dz \:\frac{z^{\,N-1} -1}{1-z} ~ \bigg[ \int_{\mu_f^2}^{(1-z) Q^2} 
  \frac{dq^2}{q^2} \, A^{\rm q}\big(\alpha_{\rm s}(q^2)\big) 
  + B^{\:\!\rm DIS}\big(\als((1-z)q^2)\big) \bigg] \:.
\end{align}	
Here $A^{\rm q}$ is the (light-like) quark cusp anomalous dimension and 
$B^{\:\!\rm DIS}$ is the resummation coefficient for DIS. Both have perturbative
series, $A^{\rm q}=\sum_i \, a_s^i \, A^{\rm q}_i$ etc, in terms of the strong 
coupling which we normalize as $\as\equiv \alpha_{\rm s}/4\pi$.
Performing the integrations one can organize the exponent as 
\begin{align}
\label{eq:GN}
G^{\,N} &= \ln N \, g^{(1)}(\lambda) + g^{(2)}(\lambda) + \as g^{(3)}(\lambda) 
  + \a2 g^{(4)}(\lambda) + \a3 g^{(5)}(\lambda) + \ldots \; ,
\end{align} 
where $\lambda = \btz \:\!\as \ln\, N$ 
or $\lambda = \btz \:\!\as \ln \,\widetilde{N}$
with $\ln \,\widetilde{N} = \ln\, N + \gamma_{\,\rm e\,}$.
The first $n\!+\!1$ terms in (\ref{eq:GN}) are required for the resummation at 
N$^n$LL accuracy. The N$^2$LL and N$^3$LL contributions to $G^{\,N}$ have been 
derived in \cite{Gnnll,Moch:2005ba};
explicit expressions can be found in (3.3) -- (3.6) of \cite{Moch:2005ba}. 
The lengthy new function $g^{(5)}(\lambda)$ entering at N$^4$LL will be 
presented in \cite{das:dis}. 
The $N$-independent prefactor $g_0^{}$ is presently known to order $\al3$ from 
the all-$N$ calculation in \cite{Cdis3},
see (4.6) -- (4.8) of \cite{Moch:2005ba},
\begin{align}
\label{eq:defg01}
 g_0^{} = 1 + \as g^{}_{01} + \a2 g^{}_{02} + \a3 g^{}_{03} 
        + {\cal O}(\a4) \,.
\end{align} 
The resummation to N$^4$LL requires the terms up to $A_5^{\rm q}$ and 
$B_4^{\:\!\rm DIS}$ in their corresponding expansions. 
The impact of the former quantity, for which a first estimate has been obtained
in \cite{Herzog:2018kwj}, is very small.
$B^{\:\!\rm DIS}$ can be calculated from knowledge of the quark form factor or
the DIS Wilson coefficients. The form factor satisfies a differential equation 
which follows from the renormalization group and gauge invariance. Its solution 
can be found in terms of the cusp anomalous dimension $A^{\rm q}$ and the 
function $G^{\rm q}$ containing the quantity $f^{\,\rm q}$ related to a 
universal eikonal anomalous dimension and the coefficient $B^{\:\!\rm q}$ of 
$\delta(1-x)$ in the quark-quark splitting function.
The four-loop coefficient of $G^{\:\!\rm q}$ (which appears in the 
$1/\epsilon$ coefficient in the solution of the form factor) can be written as
\begin{align}
  G^{\,\rm q}_4 = 2 B_4^{\,\rm q} + f_4^{\,\rm q} + \beta_2 f_{01}^{\,\rm q} 
  + \beta_1 f_{02}^{\,\rm q} + \beta_0 f_{03}^{\,\rm q} 
  + {\cal O}(\epsilon) \:,
\end{align}
where the quantities $f_{0{\rm n}}^{\,\rm q}$ are (combinations of) known
lower-order coefficients of $G^{\,\rm q}$, see \cite{Moch:2005tm} and (20) 
of \cite{Ravindran:2006cg}.
Hence $f_4^{\,\rm q}$ can be determined in the large-$n_c$ limit from the
results of \cite{Lee:2016ixa,Moch:2017uml}. We find
\bea
\label{eq:f4qLnc}
\nonumber 
f_4^{\,\rm q}\Bigl|_{{\rm L}\nc} &=& 
         \cf\*\nct \* \biggl(
            \frac{9364079}{6561}
          - \frac{1186735}{729} \* \zeta_2
          - \frac{837988}{243} \* \zeta_3
          + \frac{115801}{27} \* \zeta_4
          + \frac{11896}{9} \* \zeta_2 \* \zeta_3
          + 3952 \,\* \zeta_5
\\
&& \nonumber \mbox{}
          - \frac{4796}{9} \* \zeta_3^2
          - \frac{129547}{54} \* \zeta_6
          - 416 \,\* \zeta_2 \* \zeta_5
          - 720 \,\* \zeta_3 \* \zeta_4
          - 1700 \,\* \zeta_7
          \biggr)
       + \cf\*\ncs\*\nf \* \biggl(
          - \frac{247315}{432}
\\
&& \nonumber \mbox{}
          + \frac{412232}{729} \* \zeta_2
          + \frac{102205}{243} \* \zeta_3
          - \frac{7589}{6} \* \zeta_4
          - \frac{824}{9} \* \zeta_2 \* \zeta_3
          - \frac{740}{9} \* \zeta_5
          + \frac{2816}{9} \* \zeta_3^2
          + \frac{15611}{27} \* \zeta_6
          \biggr)
\\
&& \nonumber \mbox{\hspn}
       + \cf\*\nc\*\nfs \* \left(
            \frac{329069}{17496}
          - \frac{22447}{729} \* \zeta_2
          + \frac{25300}{243} \* \zeta_3
          + \frac{140}{3} \* \zeta_4
          - \frac{176}{9} \* \zeta_2 \* \zeta_3
          - \frac{856}{9} \* \zeta_5
          \right)
\\
&& \mbox{\hspn}
       + \cf\*\nft \* \left(
          - \frac{16160}{6561}
          - \frac{16}{81} \* \zeta_2
          - \frac{400}{243} \* \zeta_3
          + \frac{128}{27} \* \zeta_4
          \right)
\, .
\qquad\quad
\eea
The $\a4$ contribution to resummation $B^{\,\rm DIS}$ reads, in terms of the 
genuine four-loop contributions $f_4^{\,\rm q}$ and $\,B_4^{\,\rm q}$, which are
exactly known only in the L$n_c$ limit for now, and lower-order coefficients, 
\bea
\label{eq:xB4}
B_{4}^{\:\!\rm DIS}  &\! =\! & 
  - f_4^{\,\rm q} - B_4^{\,\rm q}
  - \beta_2\*\left( 
      f_{01}^{\,\rm q}
    + g_{01}
    - \frac{1}{2}\*\zeta_2\,\*A_1^{\rm q} 
  \right)
  + \beta_0^3\*\left( 
      3\*\zeta_2\*f_{01}^{\,\rm q}
    + 3\*\zeta_2\*g_{01}^{}
    + 2\*\zeta_3\*f_{1}^{\,\rm q}
    + 2\*\zeta_3\*B_1^{\,\rm q}
  \right.
 \nonumber\\[0.5mm]
&&\mbox{\hspn}
  \left.
    + \frac{3}{2}\*\zeta_4\,\*A_1^{\rm q} 
    - \frac{3}{4}\*\zeta_2^2\*A_1^{\rm q} 
  \right)
  + \beta_0\*\beta_1\*\left( 
      \frac{5}{2}\*\zeta_2\*f_{1}^{\,\rm q}
    + \frac{5}{2}\*\zeta_2\*B_1^{\,\rm q}
    + \frac{5}{3}\*\zeta_3\,\*A_1^{\rm q} 
  \right)
  + \beta_0^2\*\left( 
      3\*\zeta_2\*f_{2}^{\,\rm q}
    + 3\*\zeta_2\*B_2^{\,\rm q}
    + 2\*\zeta_3\,\*A_2^{\rm q} 
  \right)
      \nonumber\\[0.5mm]
&&\mbox{\hspn}
  - \beta_1\*\left( 
      f_{02}^{\,\rm q}
    + 2\*g_{02}^{}
    - \left(g_{01}^{}\right)^2
    - \zeta_2\,\*A_2^{\rm q} 
  \right)
  - \beta_0\*\left( 
      f_{03}^{\rm q}
    + 3\*g_{03}^{}
    - 3\*g_{02}^{}\*g_{01}^{}
    - \left(g_{01}^{}\right)^3
    - \frac{3}{2}\*\zeta_2\,\*A_3^{\rm q} 
  \right)
\:\: ,
\eea
where $g^{}_{0i}$ are to be taken without the $\gamma_{\,\sf e\,}$ terms in 
(4.6) -- (4.8) of~\cite{Moch:2005ba}. Its explicit form is given by
\bea
\label{eq:B4disLnc}
\nonumber 
 \; B_{4}^{\,\rm DIS}\Bigl|_{{\rm L}\nc} \!\! &=& 
  \cf\* \nct \* \biggl(
      -\frac{2040092429}{139968}
      +\frac{23011973}{1944}\*\zeta_2
      +\frac{517537}{36}\*\zeta_3
      -\frac{312481}{36}\*\zeta_4
      -\frac{39838}{9}\*\zeta_2\*\zeta_3
\\[-1mm]
&& \nonumber \mbox{}
      -\frac{50680}{9}\*\zeta_5
      -988\,\*\zeta_3^2
      +\frac{12467}{6}\*\zeta_6
      +496\,\*\zeta_2\*\zeta_5
      +688\,\*\zeta_3\*\zeta_4
      +2260\,\*\zeta_7
  \biggr)
\\
&& \nonumber \mbox{\hspace*{-3mm}}
   + \cf\* \ncs\*\nf \* \biggl(
       \frac{83655179}{11664}
      -\frac{5160215}{972}\*\zeta_2
      -\frac{639191}{162}\*\zeta_3
      +\frac{24856}{9}\*\zeta_4
      +\frac{8624}{9}\*\zeta_2\*\zeta_3
\\
&& \nonumber \mbox{}
      +200\,\*\zeta_5
      -32\,\*\zeta_3^2
      -\frac{1201}{3}\*\zeta_6
   \biggr)
   + \cf\* \nft \* \left(
       \frac{50558}{2187}
      +\frac{80}{81}\*\zeta_3
      -\frac{1880}{81}\*\zeta_2
      +\frac{40}{9}\*\zeta_4
   \right)
\\
&& \mbox{\hspace*{-3mm}}
   + \cf\* \nc\*\nfs \* \biggl(
      -\frac{5070943}{5832}
      +\frac{160903}{243}\*\zeta_2
      +\frac{14618}{81}\*\zeta_3
      -\frac{2110}{9}\*\zeta_4
      -\frac{400}{9}\*\zeta_2\*\zeta_3
      +\frac{904}{9}\*\zeta_5
   \biggr)
.  \qquad
\eea

\section{Numerical implications}

\noindent
The lower-order coefficients $B_{l}^{\,\rm DIS}$ have the same structure as 
(\ref{eq:xB4}), i.e., they contain $-f_l^{\,\rm q} -B_l^{\,\rm q}$ and 
lower-order coefficients. 
Therefore, by comparing the exact results to an approximation at N$^l$LL 
in which the L$n_c$ expression for $-f_l^{\,\rm q} - B_l^{\,\rm q}$ is used 
together with the exact lower-order coefficients, we can check whether 
(\ref{eq:xB4}) with the L$n_c$ results for $f_4^{\,\rm q}$ and $B_4^{\,\rm q}$
can be expected to provide a good approximation for $B_{4}^{\:\!\rm DIS}$ 
and hence $G^{\,N}$ at the N$^4$LL accuracy of full QCD. 
 
\vspace*{1mm}
This comparison is carried out in Fig.~1 for $l=2$ and $l=3$ (at $l=1$ there
is no difference between the L$n_c$ limit and full QCD). 
The L$n_c$ curves are off by less than 0.5\% at N$^2$LL and 0.25\% at N$^3$LL
for $G^{\,\rm DIS}$ in the $N$-range shown and, at $x \leq 0.9$, for the 
convolution of its exponential with a schematic but sufficiently realistic
form for a quark PDF.
Therefore we can safely expect that the L$n_c$ numbers will deviate from 
(presumably exceed) the exact QCD results by well below 1\%.

\begin{figure}[thb]
\vspace*{2mm}
\centerline{
\includegraphics[width=7.5cm, height=8.5cm]{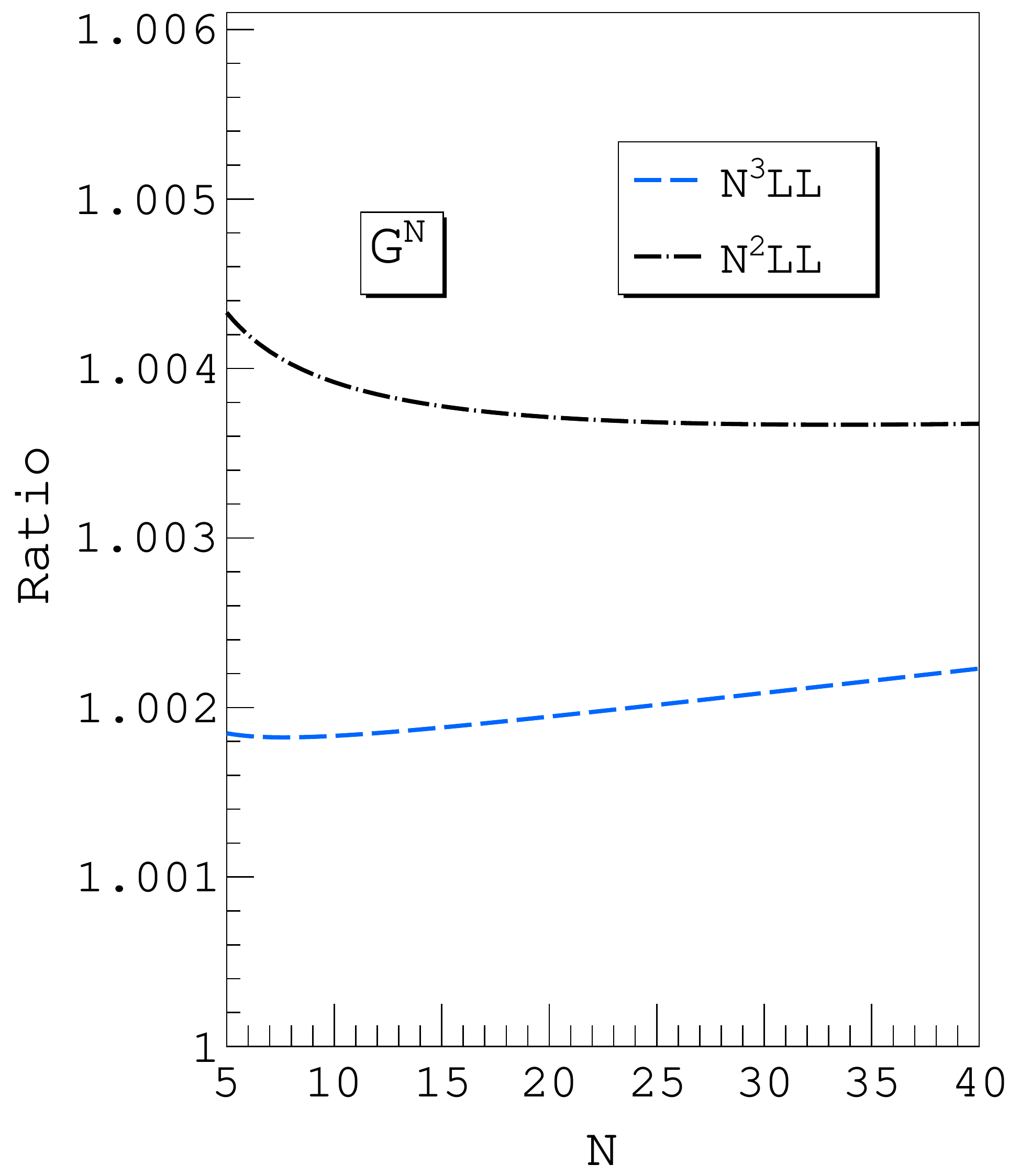}
\includegraphics[width=7.5cm, height=8.5cm]{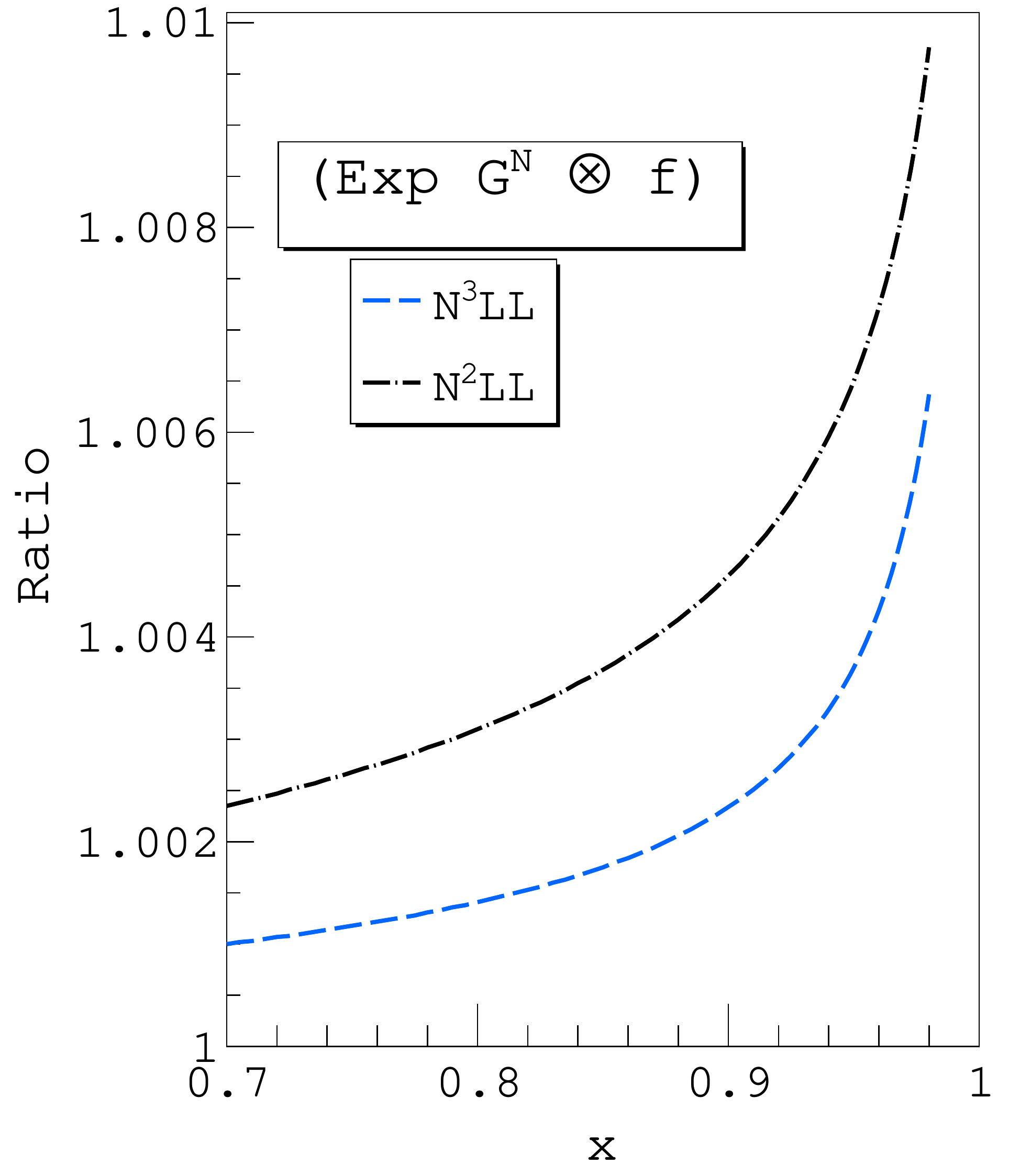}
}
\vspace*{-2mm}
\caption{
  The ratio of the large-$\nc$ approximation, defined as above, and the 
  exact results at N$^2$LL and N$^3$LL for the DIS resummation exponent 
  $G^{\,N}$ (left) and for the convolution of the exponential with 
  the schematic quark PDF shape $xf = x^{0.5}(1-x)^3$
  (right) for $\als = 0.2$ and $\nf = 3$ flavours.}
\label{Fig:GNp_ratio}
\end{figure}

The cumulative effect, relative to the NLL results, of the exact N$^2$LL and 
N$^3$LL contributions and our new N$^4$LL corrections, as above determined 
using the L$n_c$ limit of $-f_l^{\,\rm q} -B_l^{\,\rm q}$ in (\ref{eq:xB4}), 
is illustrated in Fig.~2. Unlike the N$^3$LL contribution, the N$^4$LL 
correction is almost negligible at $N \leq 15$ and $x \leq 0.9$.  
Even at $N=40$, the functions $g^{(n)}(\lambda)$ add only 6\%, 1.6\% and 1\%
to the NLL result, respectively, for $n=2$, $n=3$ and $n=4$, where the latter
L$n_c$ result is presumably a slight overestimate. The corresponding N$^2$LL,
N$^3$LL and N$^4$LL percentages for the convolution of $\exp G^{\,N}$ with 
$xf = x^{0.5} (1-x)^3$ at $x = 0.95$ read 9.5\%, 1.5\% and 0.5\%, where we 
have performed the Mellin inversion using a standard contour, see, e.g.,
\cite{Vogt:2004ns}, which constitutes a `minimal prescription' contour
\cite{Catani:1996yz} in the context of the present exponentiation. It appears
that the expansion of $G^{\,N}$ to N$^4$LL for the structure functions in 
inclusive DIS is sufficient for all practical purposes.

\begin{figure}[t]
\centerline{
\includegraphics[width=7.5cm, height=8.0cm]{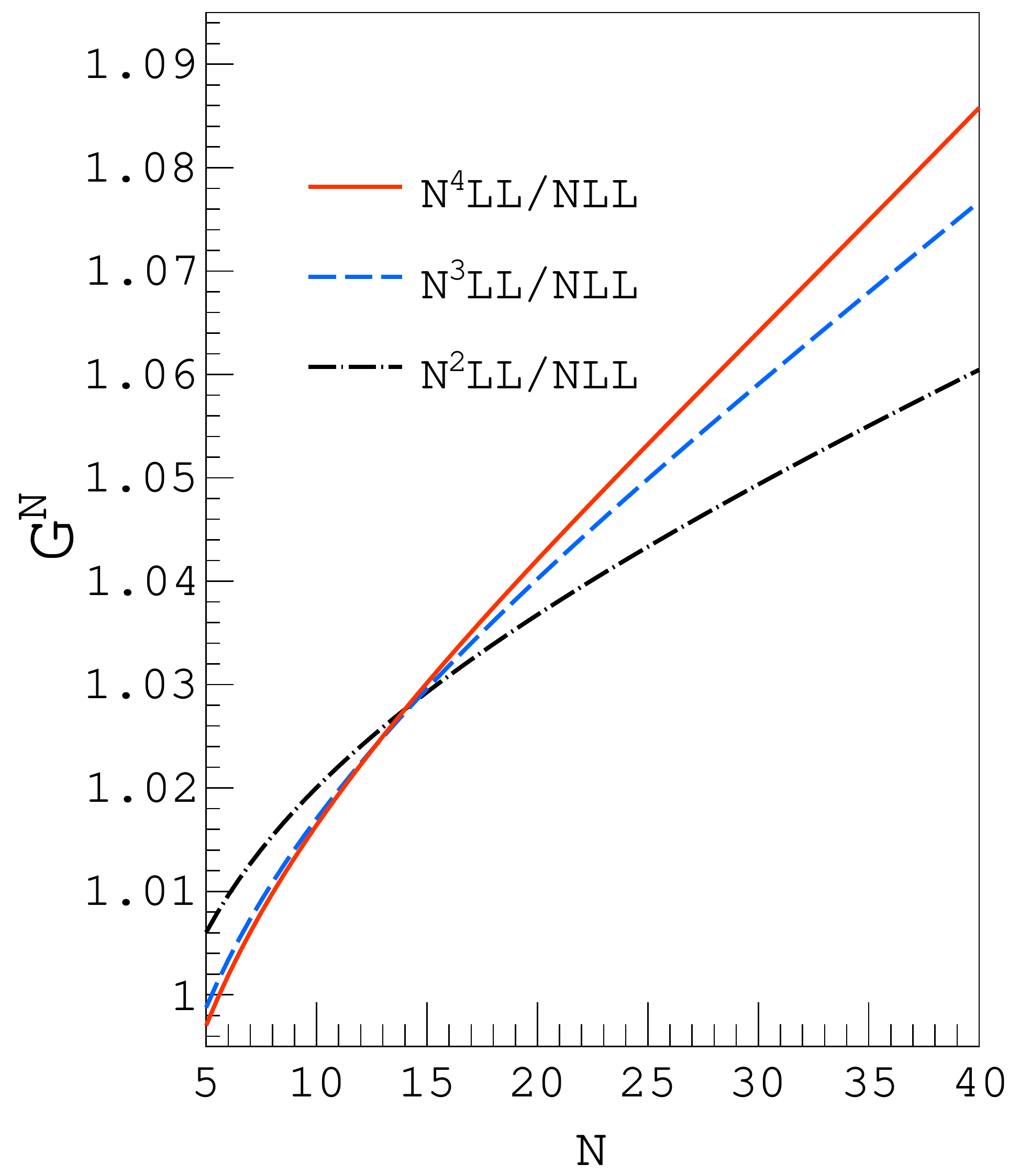}
\hspace*{-3mm}
\includegraphics[width=7.5cm, height=8.0cm]{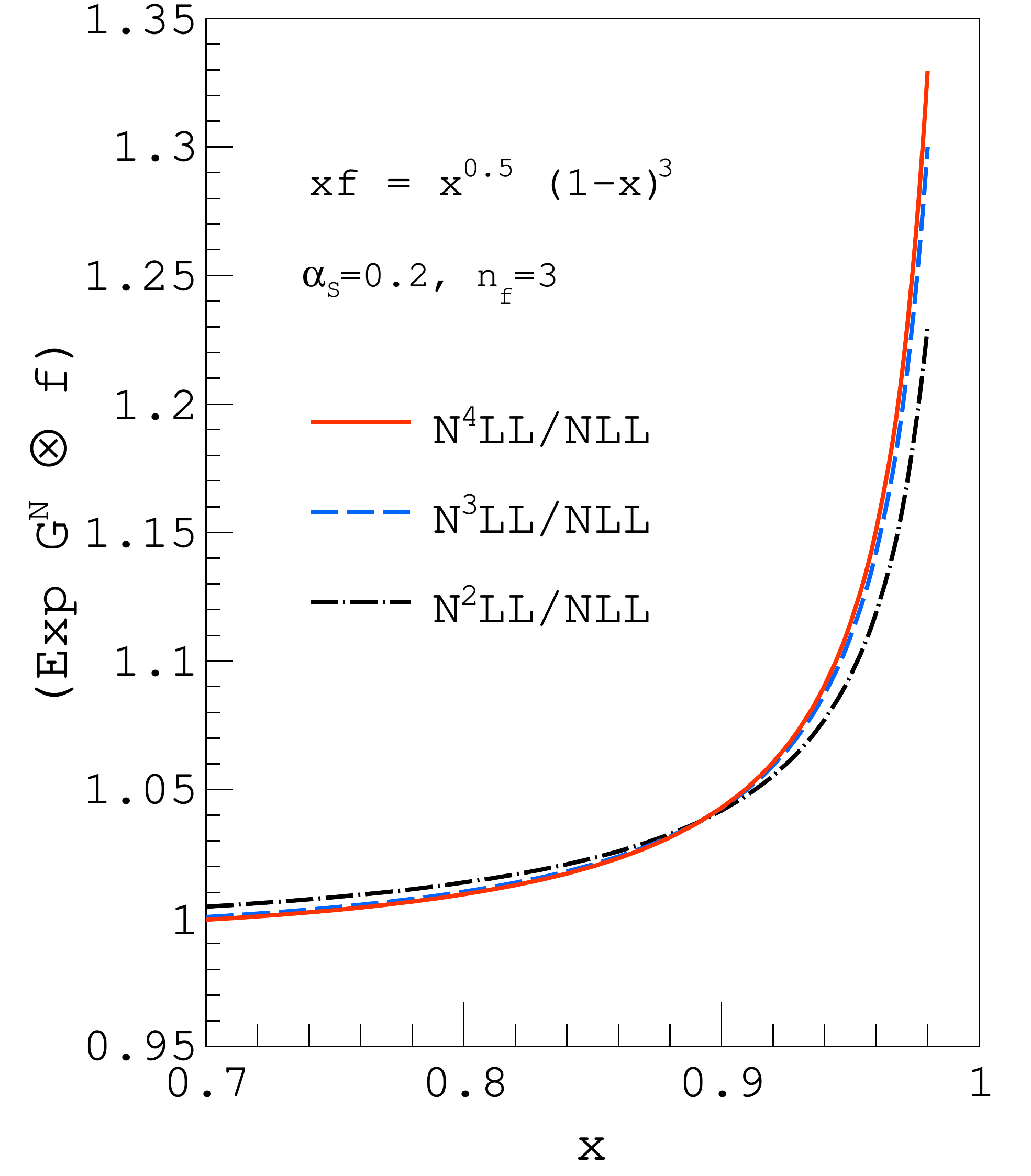}
}
\vspace*{-2mm}
\caption{
  Left: The DIS resummation exponent $G^{\,N}$ in (\ref{eq:GN}) up to 
  N$^4$LL accuracy, normalized to the NLL result at the standard 
  reference point $\als = 0.2$ for $n_f = 3$.
  Right: corresponding $x$-space results for $\exp G^{\,N}$ after convoluted 
  with a schematic form of a quark PDF of the proton.}
\label{Fig:GNp}
\vspace*{-2mm}
\end{figure}

\section{Summary and outlook}

\noindent
We have studied the soft-gluon exponentiation (SGE) of inclusive DIS at the 
fifth logarithmic (N$^4$LL) order. Recent four-loop results on splitting 
functions and the quark form factor \cite{Lee:2016ixa,Moch:2017uml} facilitate
the exact determination of the form-factor coefficient $f^{\,\rm q}$ and the
SGE coefficient $B^{\,\rm DIS}$ at order $\al4$ in the large-$n_c$ (L$n_c$)
limit. Both coefficients are relevant beyond the context of DIS:
Like the lightlike quark and gluon cusp anomalous dimensions $A^{\rm q,g}$
\cite{Pnnlo},
the quantities $f^{\,\rm q,g}$ are maximally non-Abelian and related by a
simple Casimir scaling up to three loops. We expect that the generalized
Casimir scaling of \cite{Moch:2018wjh} also applies to $f^{\,\rm q,g}$,
hence our result (\ref{eq:f4qLnc}) fixes also $f^{\,\rm g}$ at large $n_c$.
The coefficient here called $B^{\,\rm DIS}$ is due to the outgoing unobserved
quark; hence it contributes to the SGE for many other processes including, 
e.g., direct photon production \cite{Catani:1998tm}.

\vspace{1mm}
The L$n_c$ approximation to the N$^4$LL resummation exponent $G^{\,N}$ for 
inclusive DIS, defined as discussed above, is sufficiently accurate to 
demonstrate that the N$^4$LL corrections are small: they contribute well 
below 1\% over a wide range in $N$ and $x$.
As shown in \cite{Moch:2009hr}, the $1/N \ln^{\:\!\ell} N$ non-SGE 
contributions are larger; the highest four of these logarithms are currently 
known to all orders \cite{Moch:2009hr,xiDIS4} -- recall the parameter 
$\xi_{\rm DIS_4}$ unspecified in \cite{Moch:2009hr} was fixed in \cite{xiDIS4}.
We have considered the case of $\nf=3$ light flavours. In electromagnetic
and neutral-current DIS, also charm production close to threshold needs to 
be taken into account beyond the threshold for $c\bar{c}$ production, see
\cite{Kawamura:2012cr}.

\end{document}